\documentclass[a4paper]{PoS}
\usepackage{graphicx}
\usepackage{epsfig}
\usepackage{amsmath}

\title{\vspace{-2cm} {\hspace{13.05cm}\normalfont\small{DESY 11-260}}\\ \vspace{0.5cm}
       Associated production of vector gauge boson \& graviton to NLO QCD}

\ShortTitle{Graviton plus vector boson production to NLO QCD }
\author{{M. C. Kumar}\\
        Deutsches Elektronen-Synchrotron DESY, Platanenallee 6, D-15738 Zeuthen, Germany\\
        E-mail: \email{kumar.meduri@desy.de}}

\author{{Prakash Mathews}\\
        Saha Institute of Nuclear Physics, 1/AF Bidhan Nagar, Kolkata 700 064, India\\
        E-mail: \email{prakash.mathews@saha.ac.in}}

\author{{V. Ravindran}\\
        Regional Centre for Accelerator-based Particle Physics, Harish-Chandra Research Institute, Chhatnag Road, Jhunsi, Allahabad 211 019, India\\
        E-mail: \email{ravindra@hri.res.in}}

\author{\speaker{Satyajit Seth}\\
        Saha Institute of Nuclear Physics, 1/AF Bidhan Nagar, Kolkata 700 064, India\\
        E-mail: \email{satyajit.seth@saha.ac.in}}

\abstract{In this talk, we discuss the next-to-leading order (NLO) QCD corrections to the associated production of the vector 
gauge boson (Z/W$^{\pm}$) and the graviton in the large extra dimension model, namely the ADD model, at the LHC. After a brief review of the ADD 
model, we present the importance of QCD correction to these preferred processes and the impact of the QCD corrections 
on the total cross sections as well as the 
differential distributions of the gauge bosons. 
The dependence of the cross sections on the arbitrary factorization 
scale is studied and the reduction in the scale uncertainties at NLO level is shown. The ultraviolet sensitivity of the 
theoretical prediction is also presented.}

\FullConference{10th International Symposium on Radiative Corrections (Applications
of Quantum Field Theory to Phenomenology)\\
                 September 26-30, 2011\\
                 Mamallapuram, India}

\begin{document}

\section{Introduction}
To address the large hierarchy between the electroweak scale and the Planck scale remains an 
interesting and challenging task for decades. A variety of models has been proposed to address the 
hierarchy problem. All these models are subject to verification and with the 
advent of the Large Hadron Collider (LHC) with 
its unprecedented energy and luminosity, it is expected to test the TeV scale gravity models, which can give rise to 
new and interesting signals.
One such beyond Standard Model (SM) candidate is the ADD model, proposed 
by Arkani-Hamed, Dimopoulos and Dvali \cite{ADD} and one such interesting signal 
is the production of vector boson in association with a graviton
leading to missing energy.

The ADD model was the first extra dimension model 
in which the compactified dimensions could be of macroscopic 
size.  A viable mechanism to hide the extra spatial dimension, is to introduce a 3-brane 
with negligible tension and localise the Standard Model (SM) 
particles on it. Only gravity is allowed to propagate in the full $4+d$ dimensional 
space time.  
For simplicity, the extra dimensions can be assumed to be flat, of the 
same size and compactified on a $d$-dimensional torus of radius $R/(2\pi)$.
After the compactification, the scale $M_s$ of the extra dimensional theory 
is related to the Planck scale $M_p$ as:
\begin{eqnarray}
M_p^2 = C_d~M_s^{2+d} ~R^d ~,
\label{relation}
\end{eqnarray}
where $C_d= {2 ~ (4\pi)^{-{\frac{d}{2}}} / \Gamma(d/2)}$ and $R$ is the size of 
the extra dimensions.  This compactification implies that a massless graviton 
propagating in $4+d$ dimensions manifests itself as a tower of massive graviton 
modes in $4$-dimensions, with mass $ m_{\vec{n}}^2  = 4 \pi^2 \vec{n}^2/R^2$ 
where $\vec{n} = \{n_1, n_2, ...., n_d\}$ and $n_i = \{0, 1, 2, ...\}$.
Here, the zero mode corresponds to the $4$-dimensional massless graviton.
As the inverse square law of gravity has been tested down to only few 
$\mu m$ so far \cite{expt}, the size of the extra spatial dimensions in 
this model can be taken as large as this limit. 
For $M_s \sim {\cal O}(\text{TeV})$, the above limit on $R$ constrains the number 
of extra dimensions to $d \ge 2$.

In the effective theory valid below the scale $M_s$, these gravitons couple 
to the SM fields through energy momentum tensor $T^{\mu\nu}$ of the latter with the 
coupling $\kappa = \sqrt{16\pi}/M_p$, as given by 
\cite{GRW, HLZ}
\begin{eqnarray}
{\cal L}_{int} &=& - \frac{\kappa}{2} \sum_{\vec n=0}^\infty
T^{\mu\nu} (x) ~h_{\mu\nu}^{(\vec n)} (x) ~.
\label{int}
\end{eqnarray}
The Feynman rules for the above interaction
Lagrangian are given in \cite{GRW,HLZ}. 
To order $\kappa^2$, the above action allows scattering processes
involving SM fields and virtual gravitons in the intermediate state 
or real gravitons in the final state.  
In the context of collider phenomenology, 
this gives rise to a very rich and interesting signals that 
can be seen at the present LHC. The virtual exchange of 
the gravitons can lead to the deviations from the SM predictions 
whereas the real emission of the gravitons can lead to the missing 
energy signals.  Though the coupling of each graviton mode to the SM fields 
is $M_p$ suppressed, the large multiplicity of the available graviton modes 
can give rise to observable effects.  
As the size of the extra dimensions could be large in this model, the mass 
splitting i.e. $2\pi/R$ is very small and hence this summation over the graviton 
modes can be approximated to be an integral in the continuum limit, with 
the density of the graviton modes given by \cite{HLZ}
\begin{eqnarray}
\rho(m_{\vec n}) = \frac{R^d ~ m_{\vec n}^{d-2}} 
{(4\pi)^{d/2} ~ \Gamma(d/2)} ~.
\end{eqnarray}
For the real graviton production process at the collider experiments, 
the inclusive cross section is given by the following convolution:
\begin{eqnarray}
{d \sigma}&=& 
\int dm_{\vec{n}}^2 ~ \rho(m_{\vec{n}}) ~ 
{d \sigma_{m_{\vec{n}}}} ~,
\label{d_state2}
\end{eqnarray}
where $d\sigma_{m_{\vec{n}}}$ is the cross section for the production of a 
single graviton of mass $m_{\vec{n}}$.

At the hadron colliders like LHC or Tevatron, the QCD radiative 
corrections are very significant for they can enhance the LO 
predictions as well as decrease the arbitrary scale uncertainties in 
theoretical predictions. Further, the presence of a hard jet in the final 
state, due to these radiative corrections, has the potential to modify 
the shapes of the transverse momentum distributions of the particles that 
are under study at LO.  Obtaining such a modification to the shapes of 
the distributions is beyond the scope of the normalization of the corresponding 
LO distributions  by a constant K-factor, and it requires an explicit computation 
of the cross sections or distributions to next-to-leading order (NLO) in QCD. 
In the context of missing energy signals in the large extra dimensional model,
the NLO QCD corrections are presented for the processes (i) jet plus graviton
production \cite{KKLZ} and (ii) photon plus graviton production \cite{GLGW}. 
In each of these two cases, it is shown that the K-factors can be as high as 
$1.5$ at the LHC.

\section{Importance of graviton plus vector boson production}
The gravitons when produced at the collider experiments escape 
the experimental detection due to their small couplings and negligible 
decays into SM particles.  The production of vector bosons ($V=Z,~W^{\pm}$) 
together with such an {\it invisible} gravitons ($G$) can give rise to a very large 
missing transverse momentum signals at the collider experiments.
The study of graviton plus gauge boson production, hence, in general will be a 
useful one in probing the new physics at the LHC. This process has been studied
at leading order (LO) in the context of lepton colliders \cite{Kingman, Giudice}
as well as at the hadron colliders \cite{Ask}, and also has been implemented 
in Pythia8 \cite{Ask2}. The process is an important one and stands complementary
to the more conventional ones involving the graviton production, like jet plus
graviton or photon plus graviton productions, that are generally useful in
the search of the extra dimensions at collider experiments.

   It is important to note that there is a Standard Model (SM) background 
which gives signals similar to those of associated production of $Z$ and
$G$.  This SM background receives a dominant contribution coming from 
the $ZZ$ production process, where one of the $Z$-bosons in the final 
state decays into a pair of neutrinos ($Z \to \nu \bar{\nu}$) leading 
to $Z$-boson plus missing energy signals. The other $Z$-boson  can be 
identified via its decays to leptons, mostly electrons and muons, and 
then constraining the lepton invariant mass close to the mass of the 
$Z$-boson to consider only the on-shell $Z$-bosons. A detailed study of 
the event selection and the minimization of other SM contributions to
this process $ZZ \to l \bar{l}\nu\bar{\nu}$, using MC@NLO and Pythia, 
is taken up in the context of ATLAS detector simulation and is 
presented in \cite{ATLAS}.  Any deviation from this SM prediction will
hint some beyond SM scenario and hence a study of this process will be
useful in searching the new physics.

In what follows, we describe the computation of NLO cross sections for the
process under study. Since our focus is on the QCD part in this work, we 
will confine our calculation to the production of on-shell $Z/W^{\pm}$.

\section{Calculational details}
At the lowest order in the perturbation theory, the associated production
of the vector gauge boson and the graviton takes place via the quark 
anti-quark initiated subprocess, given by
\begin{eqnarray}
q_a(p_1) + {\bar{q}_b}(p_2) \rightarrow V(p_3) + G(p_4) ~,
\end{eqnarray}
where $V = Z, W^\pm$ and $a,b$ are flavor indices. 
The Feynman rules and the summation of polarization tensor of 
the graviton are given in  \cite{GRW,HLZ}. For the vector gauge 
boson, the propagator in the unitary gauge $(\xi \rightarrow \infty)$ 
has been used throughout because of some advantages \cite{ss1,ss2}. 

At the NLO in the perturbation theory, the cross sections
receive ${\cal O}(\alpha_s)$ contributions from real emission as well as
virtual diagrams.  The integration over the phase space of the real emission diagrams will 
give rise to infra-red (IR) (soft and collinear) divergences in the limit 
where the additional parton at NLO is either soft and/or collinear to the initial 
state partons. On the other hand, the integration over the loop momenta in 
the virtual diagrams will also give rise to infrared divergences, in addition
to the ultraviolet (UV) divergences. In our calculation, we regulate all these 
divergences using dimensional regularization with the 
number of space-time dimensions $n = (4 + \epsilon)$. 
Completely anti-commuting $\gamma_5$ prescription \cite{CFH} is used to handle 
$\gamma_5$ in $n$ dimensions.
Here, it should be noted that as the gravitons couple to the energy momentum tensor 
of the SM fields, which is a conserved quantity, there won't be any UV divergences 
coming from the loop diagrams.  

There are several methods available in the literature to compute NLO QCD corrections.
Standard methods based on fully analytical computation deal with the phase space 
and loop integrals in $n$-dimensions and give a finite ${\cal O}(\alpha_s)$ 
contribution to the cross sections, after the real and the virtual contributions 
are added together and the initial state collinear singularities are absorbed into
the bare parton distribution functions.
However, these methods are not useful whenever the 
particles in the final state are subjected to 
experimental 
cuts or some isolation algorithms. In such cases,  semi analytical methods like 
{\it phase space slicing method} or {\it dipole subtraction method} are extremely 
useful. In the present work, we have resorted to the former \cite{Harris} with two cut offs ($\delta_s,\delta_c$) to compute the radiative corrections. 
In this method, the IR 
divergences appearing in the real diagrams can be handled in a convenient way by 
slicing the soft and collinear divergent regions from the full three body phase space. 
The advantage of this method is that the integration over the remaining 
phase space can be carried out in $4$-dimensions, rather than in $n$-dimensions, 
using standard Monte-Carlo techniques.  
For any further calculational details about the real and virtual parts of these processes, we 
refer to \cite{ss1,ss2}.

\section{Numerical Results}
In this section, we present various kinematic distributions for the 
associate production of the graviton and the  vector gauge boson to 
NLO in QCD at the LHC. The results are presented for proton-proton 
collision energy of $\sqrt{S} = 14$ TeV.  
The limits on the integral over the graviton mass are set by the kinematics from $0$ to 
$\sqrt{s} -m_V$, where $\sqrt{s}$  is the parton center of mass 
energy and $m_V = m_Z, m_W$.  
The masses of the gauge bosons and the weak mixing angle are given 
by \cite{PDG1}, $m_Z = 91.1876$ GeV, $m_W = 80.398$ GeV, $\text{sin}^2\theta_w = 0.2312$.
For $W$ boson production cross sections, we will consider 
the mixing of quarks among different quark generations, as allowed by the 
CKM-matrix elements $V_{ij}$, with $(i = u, c, t)$ and $(j = d, s, b)$. 
Since all our calculations are done in the massless limit of the partons, 
we have not included the top quark contribution in our calculation and 
set all $V_{tj}$'s to zero. 
The fine structure constant is taken to be $\alpha = 1/128$. 
Throughout our study, we have used CTEQ6L1 and CTEQ6.6M parton density sets
for LO and NLO cross sections respectively.  The strong coupling constant is 
calculated at two loop order in the $\overline{MS}$ scheme with 
$\alpha_s(m_Z) = 0.118$ ($\Lambda_{\text{QCD}} = 0.226$ GeV). 
We have also set the number of light flavors $n_f = 5$.   
The following cuts are used for our numerical study,
\begin{eqnarray}
p_{T}^{Z, W} > p_T^{min},  \quad p_{T}^{miss} >  p_T^{min},
\quad |y^{Z,W}| \le 2.5 ~.
\label{cuts1}
\end{eqnarray}
For the 2-body process, the missing transverse momentum is same as that
of the  gauge boson.  On the other hand, for the 3-body process, it need not be so 
due to the presence of an observable jet in the final state and hence 
it amounts purely to the graviton transverse momentum. The observable
jet is defined as the one that satisfies the following conditions:
\begin{eqnarray}
p_T^{jet} > 20 \text{GeV} \quad \text{and} \quad |\eta^{jet}| \le 2.5 ~.
\label{cuts2}
\end{eqnarray}
Whenever the jet does not satisfy the above conditions, the missing 
transverse momentum is approximated to be that of the gauge boson.

\begin{figure}[htb]
\centerline{
\epsfig{file=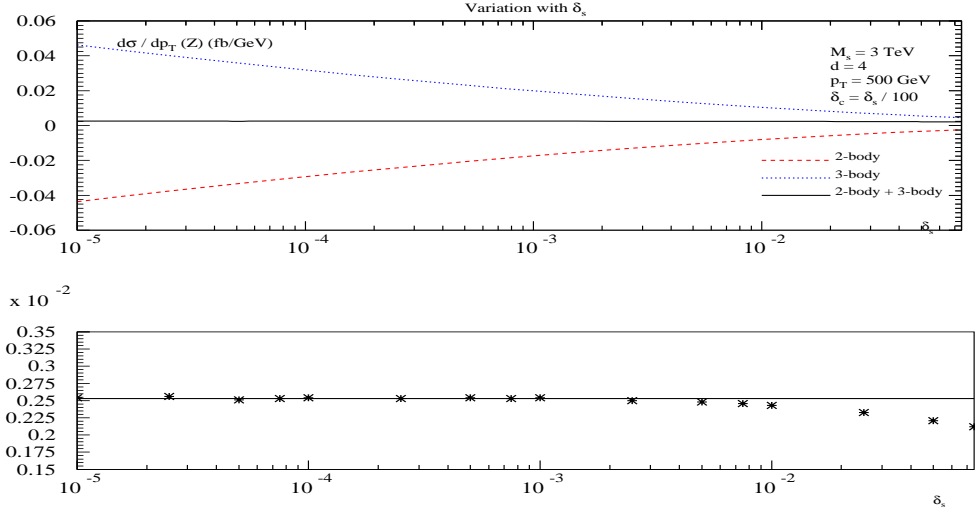,width=14cm,height=7.952cm,angle=0}}
\caption{Variation of the transverse momentum distribution of $Z$ boson
with $\delta_s$, keeping the ratio $\delta_s/\delta_c=100$ 
fixed, for $M_s=3$ TeV and $d=4$.}
\label{ds-z}
\end{figure}

\begin{figure}[htb]
\centerline{
\epsfig{file=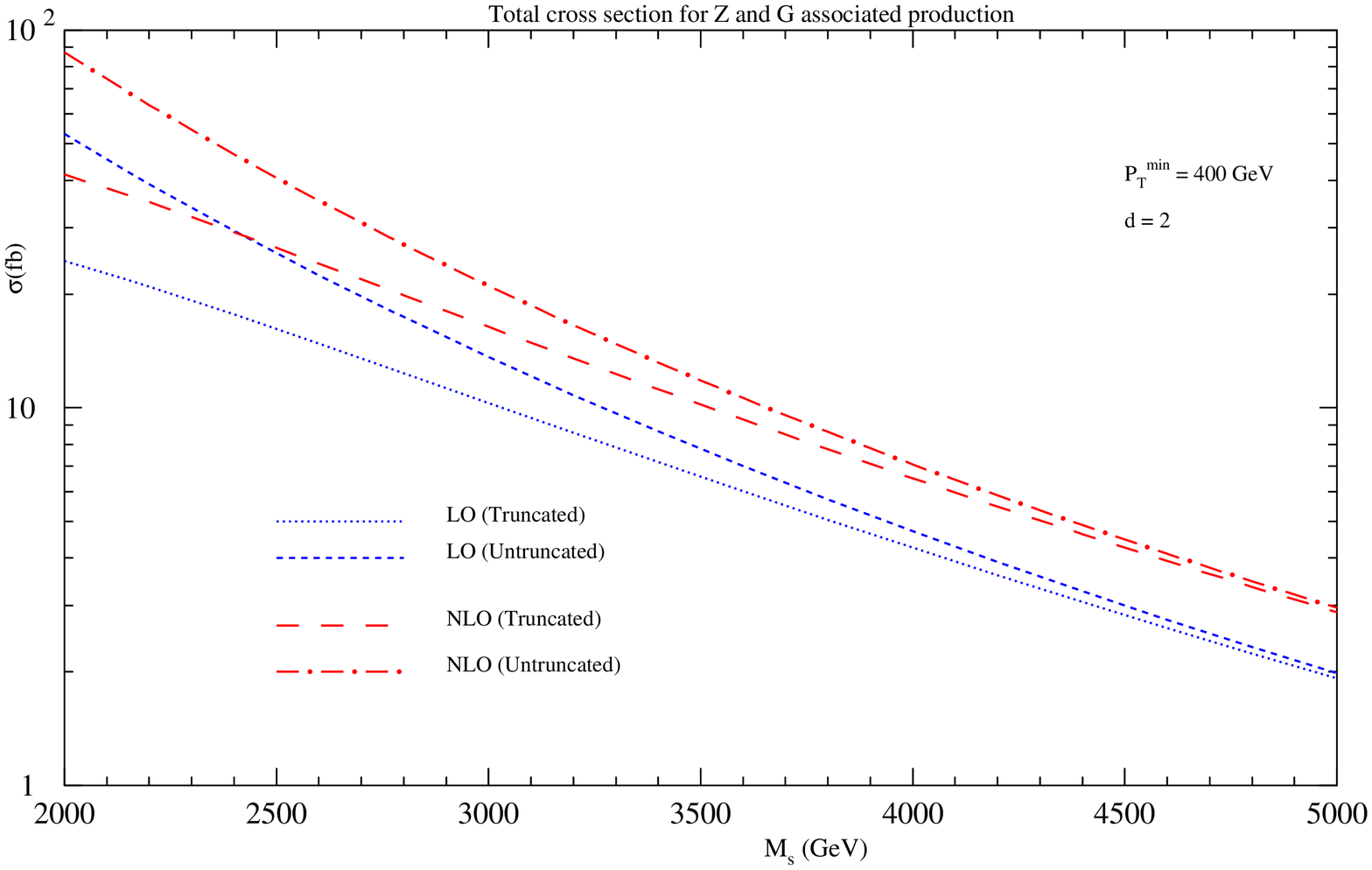,width=7.5cm,height=7cm,angle=0}
\epsfig{file=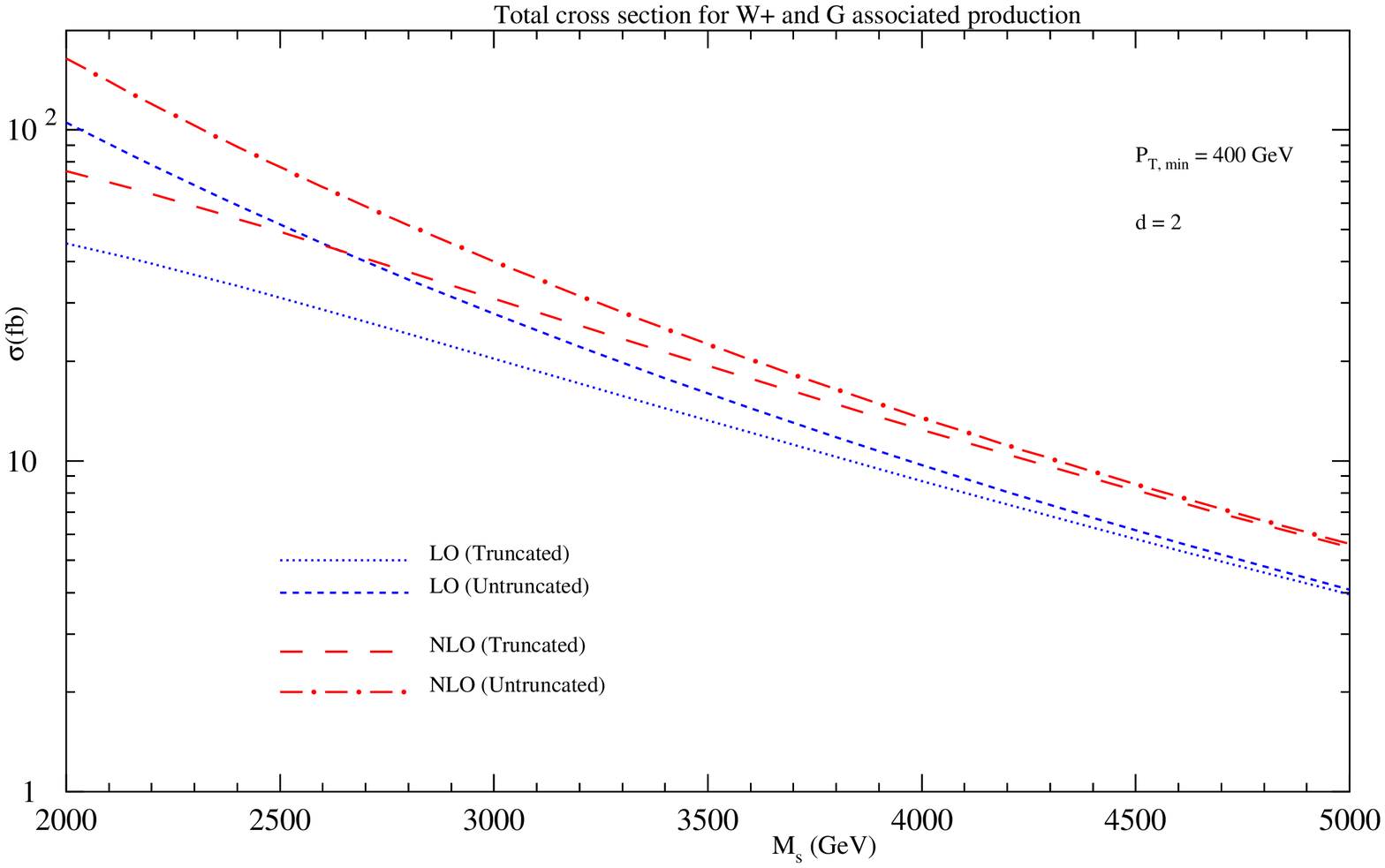,width=7.5cm,height=7cm,angle=0}}
\caption{Total cross section for the associated production of $ZG$ (left) and $W^+G$ (right)
at the LHC, shown as a function of $M_s$ for $d=2$.}
\label{totms-zwp-d2}
\end{figure}

\begin{figure}[htb]
\centerline{
\epsfig{file=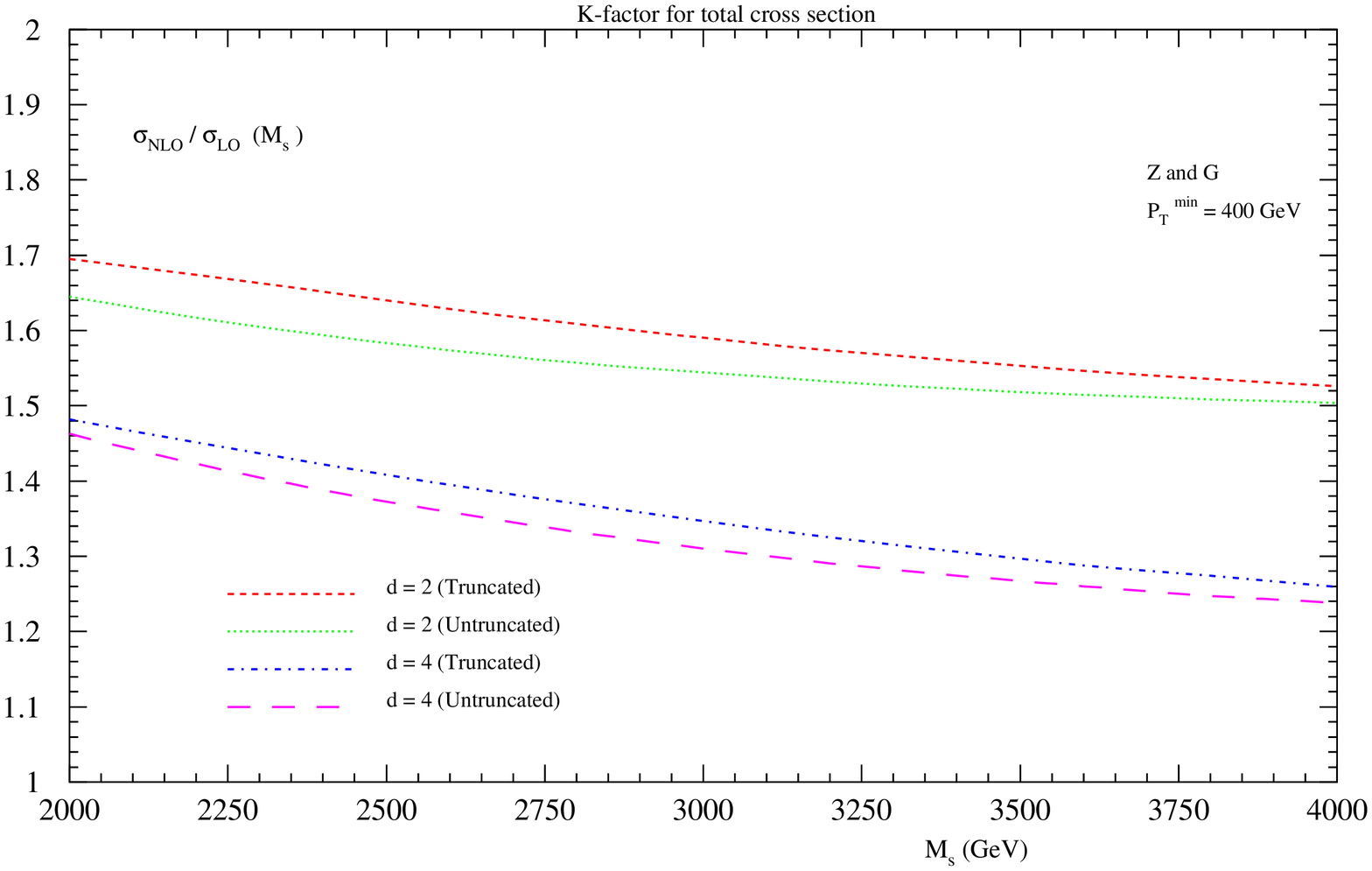,width=7.5cm,height=7cm,angle=0}
\epsfig{file=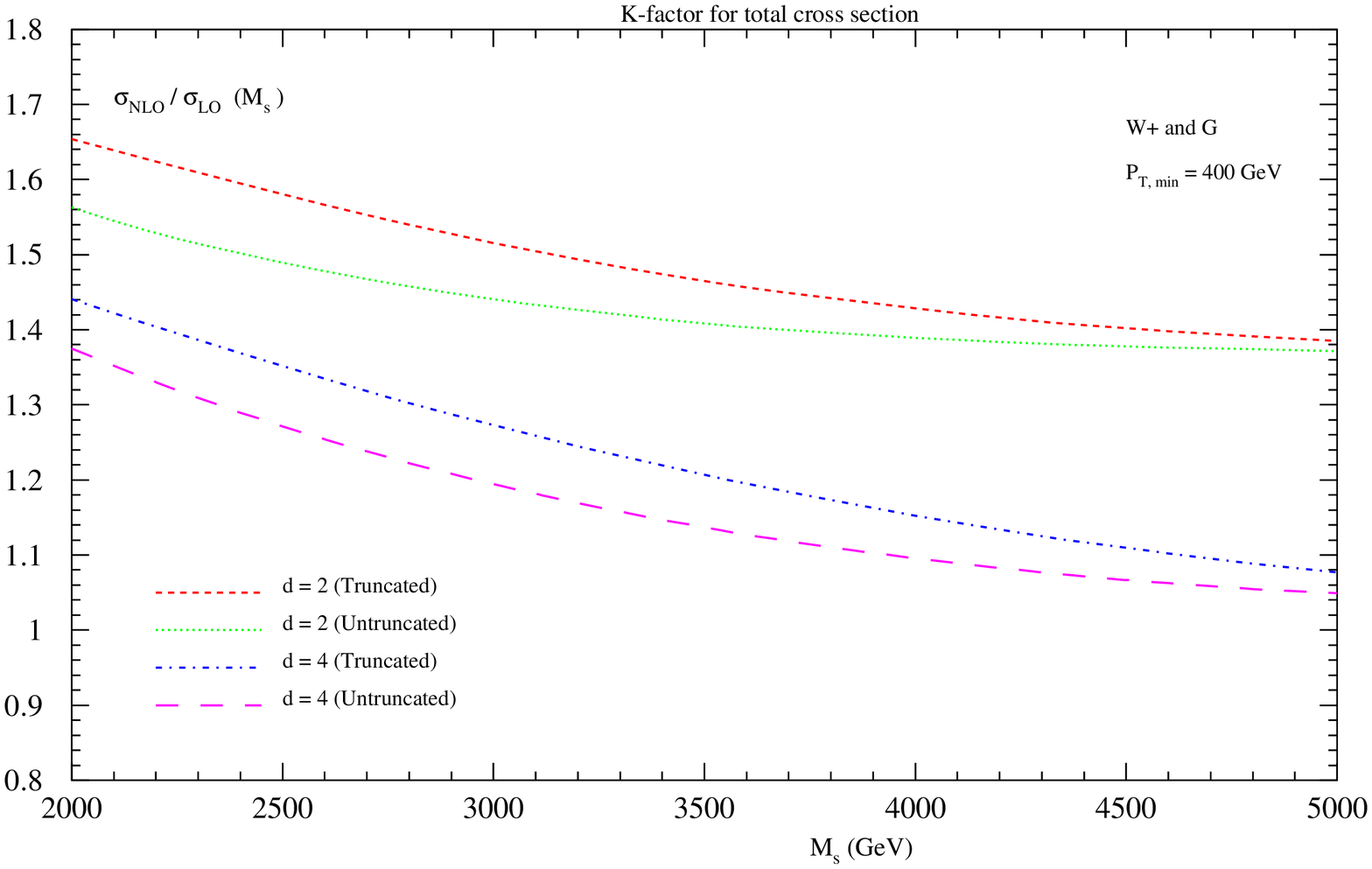,width=7.5cm,height=7cm,angle=0}}
\caption{K-factors of the total cross section for the associated production 
of the $Z$-boson  and the graviton at the LHC, given as a function of the scale $M_s$ for $ZG$ (left) 
and $W^+G$ (right) production.}
\label{tot-zwmwp-kf}
\end{figure}

\begin{figure}[htb]
\centerline{
\epsfig{file=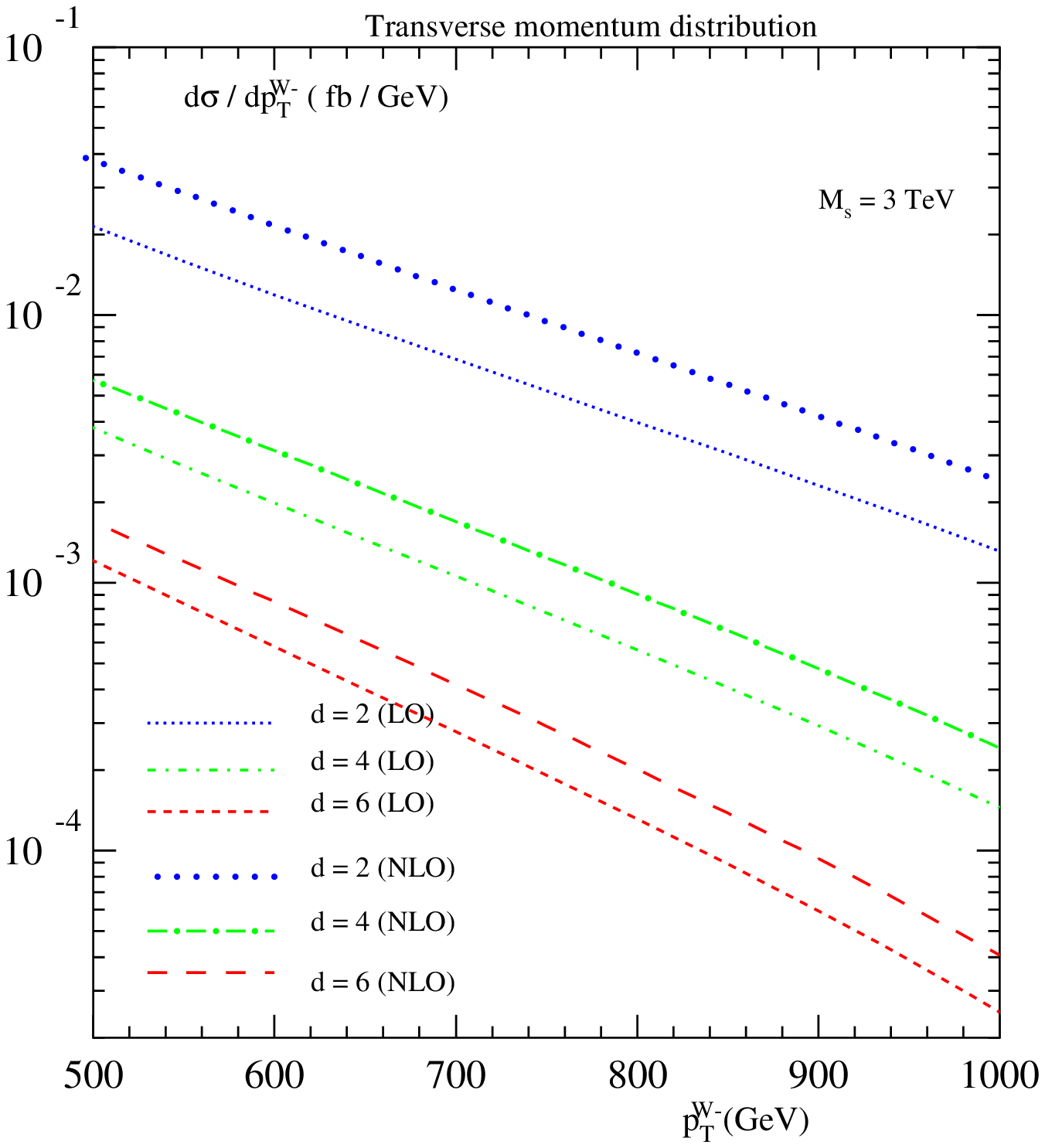,width=7cm,height=7cm,angle=0}
\epsfig{file=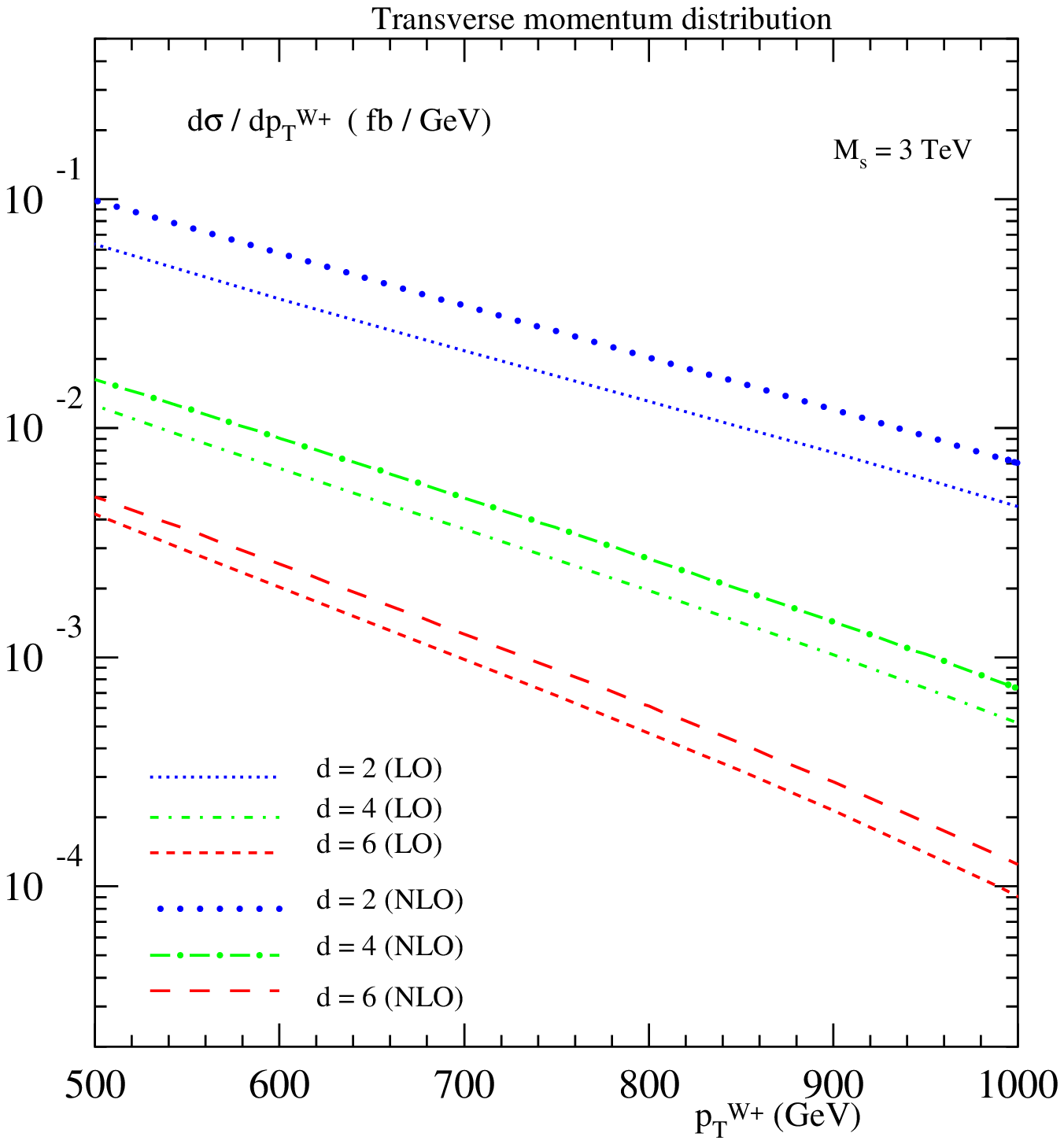,width=7cm,height=7cm,angle=0}}
\caption{Transverse momentum distribution of the $W^-$ (left) and $W^+$ (right) for $M_s = 3$ 
TeV is shown for different values of the number of extra dimensions $d$.}
\label{pt-wmwp}
\end{figure}

\begin{figure}[htb]
\centerline{
\epsfig{file=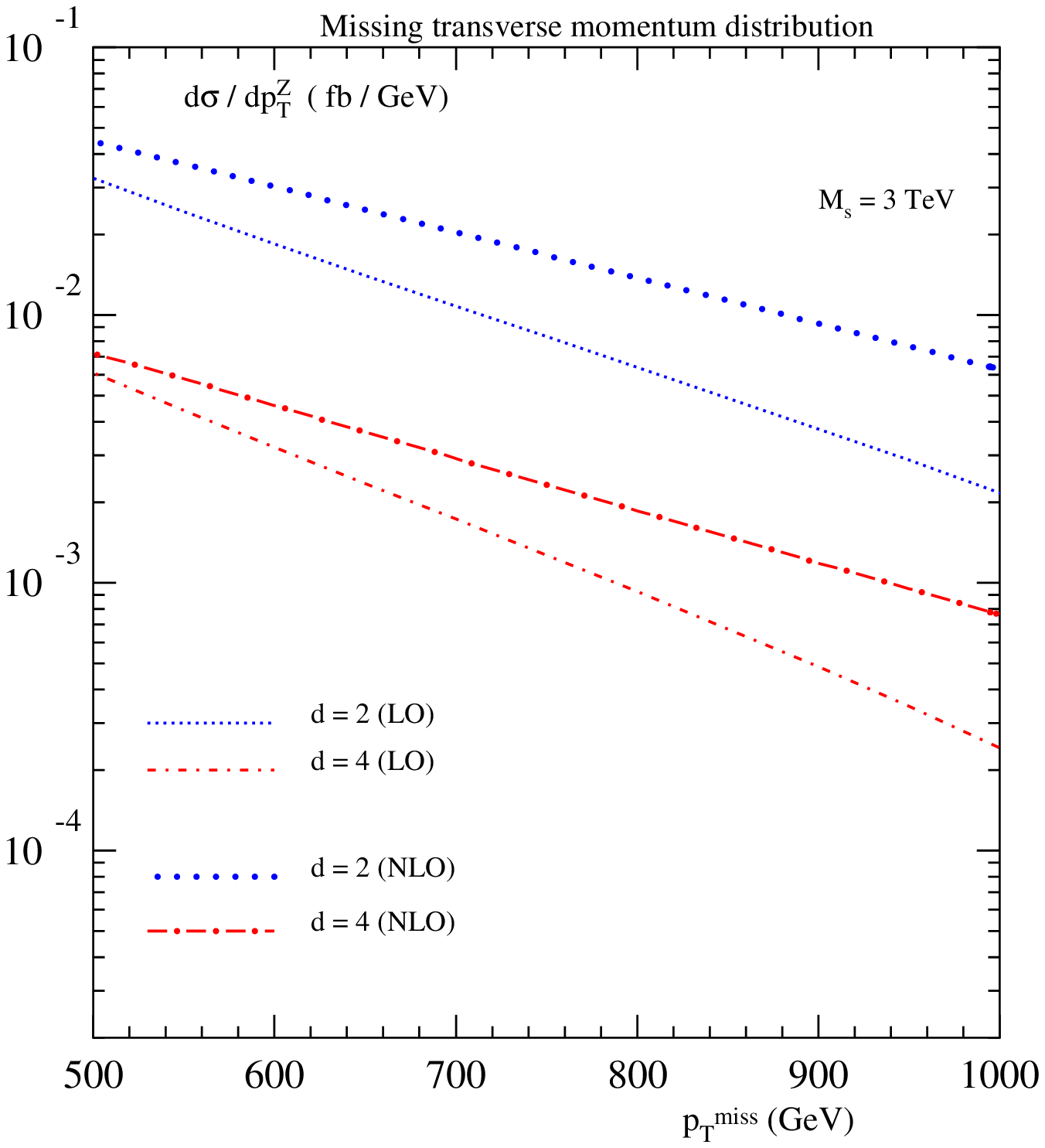,width=7cm,height=7cm,angle=0}
\epsfig{file=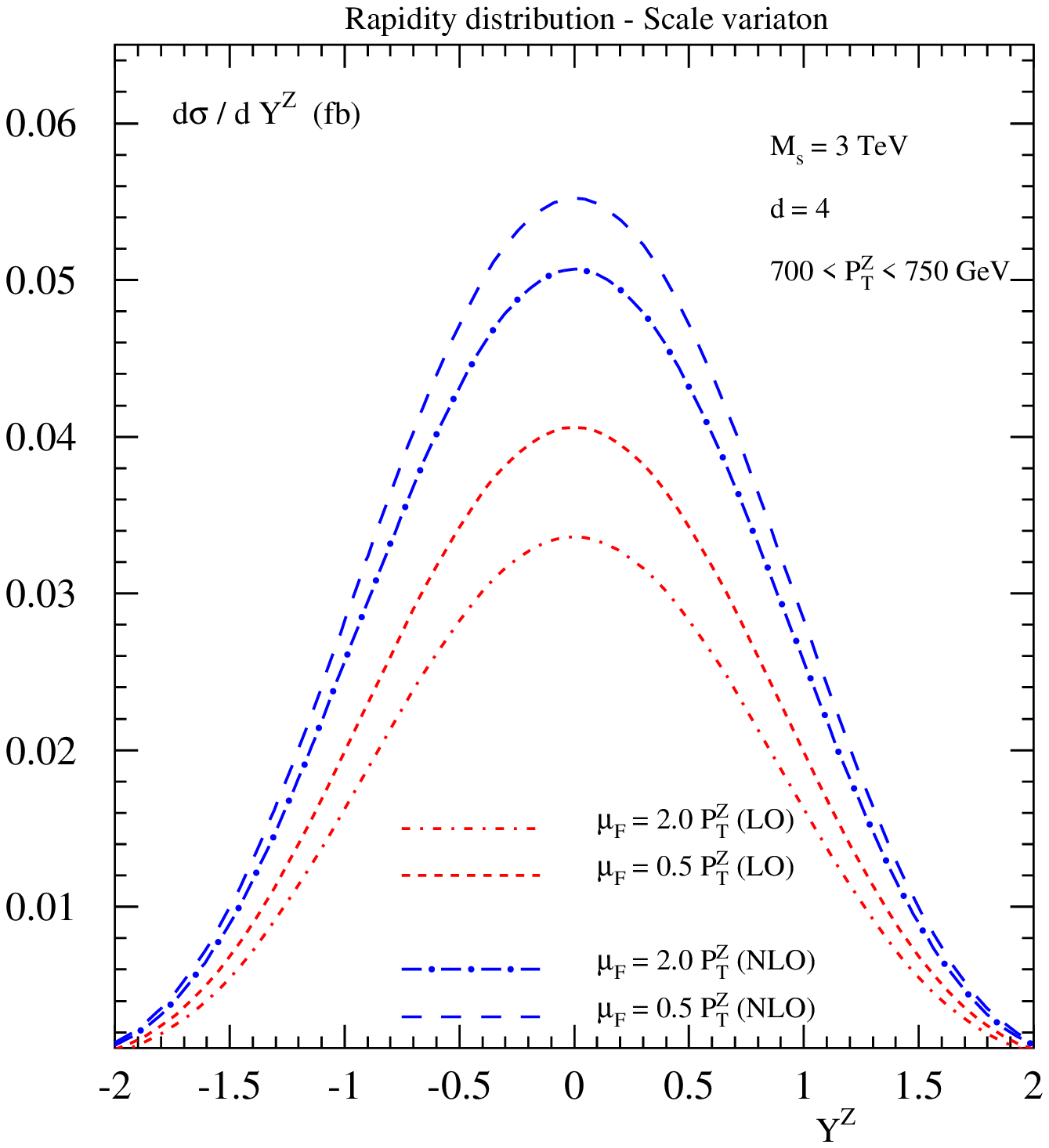,width=7cm,height=7cm,angle=0}}
\caption{Missing transverse momentum distribution of the graviton produced
in association with $Z$-boson at the LHC, for $M_s = 3$ TeV (left). The
scale uncertainties in the rapidity distribution of $Z$-boson for 
$M_s = 3$ TeV and $d=4$ (right).}
\label{ptmiss-rap-z}
\end{figure}

We check for the stability of the cross 
sections against the variation of the slicing parameters, $\delta_s,\delta_c$ for all of the three processes 
and find that our results are independent of the choice of these slicing parameters that are introduced 
in the intermediate stages of the calculation. It can be seen from the fig.\ (\ref{ds-z}) that 
both the 2-body and the 3-body contributions vary with $\delta_s$ but their sum is fairly stable 
against the variation of $\delta_s$ over a wide range. 
The cross sections are given for both the truncated as 
well as the un-truncated cases, as the ADD model is an effective theory \cite{GRW}, with the choice of model parameters 
$M_s = 3 $ TeV and $d =2$. The same is plotted for $W^+$ case \cite{ss2} also. All of these plots are studied with different 
value of $d$. In fig.\ (\ref{totms-zwp-d2}), we have shown the 
variation of the truncated as well as un-truncated total cross sections 
with respect to the scale $M_s$, for the case $d = 2$ for the $Z/W^+$ and the graviton associated production. 
The K-factors are shown in fig.\ (\ref{tot-zwmwp-kf}) for $Z$ (left panel) and $W^+$ (right panel) production with 
the graviton as a function of $M_s$. A similar study of variation of K-factors with $P_T^{min}$ for all of these three cases 
is done \cite{ss2}.
Further, in fig.\ (\ref{pt-wmwp}), we present the transverse momentum 
distribution of $W^-$ (left panel) and $W^+$ (right panel) respectively as a function of the number of extra 
dimensions $d$ and for $M_s = 3$ TeV. The missing transverse momentum distributions ($P_T^{miss}$) for the $Z$-boson case is plotted 
in the left panel of fig.\ (\ref{ptmiss-rap-z}) for $d=2, 4$. In the right panel of fig.\ (\ref{ptmiss-rap-z}), the rapidity distribution of the $Z$-boson 
both at LO and at NLO for two different choices of the factorization scale: 
$\mu_F=P_T^Z/2~~ {\rm and}~~2P_T^Z$ is plotted.  This distribution is obtained by  
integrating over the transverse momentum of the $Z$-boson from $700$ GeV 
to $750$ GeV, for $d=4$.  As expected, the inclusion of order $\alpha_s$ corrections reduces 
the dependence on the arbitrary factorisation scale $\mu_F$. 
The percentage of uncertainty in the cross sections at the central rapidity region 
$Y=0$, due the variation of the scale from $\mu_F=P_T^Z/2$ to 
$\mu_F=2 P_T^Z$, is $18.9\%$ at LO and it gets reduced to $8.6\%$ at NLO. Similar considerable amount of scale uncertainty reductions at NLO 
are observed for the rest of the two cases \cite{ss2}.  

\section{Conclusion}
We have systematically computed the full NLO QCD corrections 
to the associated production of the vector gauge boson and the graviton in theories 
with large extra dimensions at the LHC. The K-factors for the neutral gauge boson 
are found to vary from 
$1.6$ to $1.2$ depending on the number of extra dimensions $d$, while they 
vary from $1.8$ to $1.3$ for the case of charged gauge bosons. At the hadron colliders, 
the leading order predictions often suffer from large uncertainties resulting from
the choice of factorisation scale.   Reducing these uncertainties 
is one of the main motivations for doing NLO computation.  We have shown that 
this is indeed the case for the rapidity distributions of the gauge bosons 
by varying the factorization scale from $\mu_F = P_T/2$ to $\mu_F = 2P_T$, 
leading to reduction in the percentage of scale uncertainty to about $9$\% from $19$\%.  
Hence, the results presented in this paper are more suitable for 
studies on associated production of vector boson and graviton 
in the context of extra dimension searches at the hadron colliders.

\end{document}